\pdfoutput=1

\documentclass[aps, prb, amsmath, amssymb, amsfonts, twocolumn]{revtex4-1}

\usepackage{graphicx}
\usepackage{dcolumn}
\usepackage{bm}
\usepackage{amsmath}
\usepackage{comment}
\usepackage[T1]{fontenc}

\begin{document}

\preprint{APS/123-QED}

\title{Magnetic Noise from Metal Objects near Qubit Arrays}

\author{Jonathan Kenny}
\affiliation{School of Physical and Mathematical Science,\linebreak Nanyang Technological University, 21 Nanyang Link, 04-01, Singapore 637371}
 \altaffiliation[Also at ]{Physics Department, University of Wisconsin-Madison, 1150 University Ave, Madison, WI, 53706, USA}

\author{Hruday Mallubhotla}
\affiliation{Physics Department, University of Wisconsin-Madison, 1150 University Ave, Madison, WI, 53706, USA}

\author{Robert Joynt}
 \email{rjjoynt@wisc.edu}
\affiliation{Physics Department, University of Wisconsin-Madison, 1150 University Ave, Madison, WI, 53706, USA}

\date{\today}

\begin{abstract}
All metal objects support fluctuating currents that are responsible for evanescent-wave Johnson noise in their vicinity due both to thermal and quantum effects.  The noise fields can decohere qubits in their neighborhood.  It is quantified by the average value of $\bold{B}(\bold{x},t) \, \bold{B}(\bold{x'},t')$ and its time Fourier transform.  We develop the formalism particularly for objects whose dimensions are small compared with the skin depth, which is the appropriate regime for nanoscale devices.  This leads to a general and surprisingly simple formula for the noise correlation function of an object of arbitrary shape.  This formula has a clear physical interpretation in terms of induced currents in the object.  It can also be the basis for straightforward numerical evaluation.  For a sphere, a solution is given in closed form in terms of a generalized multipole expansion.  Plots of the solution illustrate the physical principles involved.  We give examples of how the spatial pattern of noise can affect quantum information processing in nearby qubits.  The theory implies that if the qubit system is miniaturized to a scale $D$, then decoherence rates of qubits scale as $1/D$.       
\end{abstract}

\maketitle


\section{Introduction}
\label{sec:introduction}


The success of quantum computing depends on being able to perform many operations before the qubits decohere.  To make the decoherence time as long as possible, we need to have a precise understanding of noise in the system.  Many platforms, particularly spin qubits and superconducting qubits that use a flux degree of freedom,  are vulnerable to magnetic noise: random fluctuations in the ambient magnetic field.  At the single qubit level, the fluctuations occur at a single point or a single small volume.  We will call these local noise correlations.  But correlations in the noise field at different spatial points are also important.  Indeed they pose dangers that are far more difficult to correct using standard quantum error correction \cite{preskill2012sufficient}. Thus it is also important to compute these nonlocal noise correlations.    

All metal objects have free random currents that create magnetic noise in the vicinity, often called evanescent-wave Johnson noise (EWJN).  This effect has been known for decades and the underlying quantum field theory was worked out in the 1960s and 1970s \cite{lifshitz, rytov,agarwal}.  The effect of electric noise from a metal surface on atom qubits was described by Henkel and collaborators \cite{henkel}, and was subsequently observed \cite{cornell}. Systematic studies of the effect of magnetic EWJN from a silver film on NV-center qubits have been performed \cite{kolkowitz,ariyaratne}. In experiments on spin qubits in Si/SiGe platforms EWJN may in many cases be responsible for the spin relaxation \cite{jiang,morello,premakumar}.

Essentially all qubit systems that would be sensitive to magnetic EWJN operate with metallic device elements.  The accidental presence of metallic inclusions is also possible.  In all cases we need to understand the strength and spatial pattern of the noise.  An obvious place to start is the noise that comes from a localized metallic object in the parameter regimes appropriate to qubit systems.  This paper presents the general theory of this problem.  The solution of a sphere in the dipole approximation is known \cite{landau,premakumar}.  However, this solution is not quantitatively correct when the distance of the object from the qubit is comparable to the size of the object, which is the parameter range that is often of interest in quantum computers. 

In this paper we solve the full problem of the sphere in a type of multipole expansion.  We also present a simple formula that can serve as the basis of a straightforward numerical solution for the noise field of an object with an arbitrary shape.
We present pictures of the noise field in order to develop some physical intuition about a physical phenomenon that is important but not necessarily widely understood by all workers in the field.  We present examples of how the noise can affect quantum information processing.

In Sec. \ref{sec:method} we set out the basic formalism, including the simplifications that are characteristic in the regimes of frequency and particle size that are relevant for nanodevices.  This allows us to derive the formula for an object of arbitrary shape.  In Sec. \ref{sec:sphere} we get the multipole expansion for the sphere. In Sec. \ref{sec:lncf} we give results for the local noise correlation function, while in Sec. \ref{sec:nncf} the results for the non-local case are presented.  Sec. \ref{sec:conclusion} is devoted to implications for experiments and prospects for future work.   

\section{Calculation Method}
\label{sec:method}

In this section we set up the formalism for the calculation, including the simplifications that are appropriate for typical quantum computing hardware.

\subsection{Noise Correlation Function}
\label{subsec:ncf}
Metal objects in the vicinity of magnetic qubits will decohere them because of the magnetic noise field set up by the random currents in the object.  We quantify this in the following way.  We choose the center of gravity of the object as the origin and compute the noise correlation function (NCF)
\begin{equation*}
\begin{split}
    \langle B_i(\vec{x}) B_j(\vec{x}')\rangle_\omega = \int^{\infty}_{-\infty} \langle B_i(\vec{x},t) B_j(\vec{x}',0)\rangle e^{i \omega t} dt.
\end{split}
\end{equation*}  
Here $B_i(\vec{x},t)$ is the $i$th Cartesian component of the magnetic noise field and the angle brackets denote a quantum and thermal average.  Both $\vec{x}$ and $\vec{x}'$ lie outside the object.  

Thermal quantum field theory has been used to work out the equation for the photon Green's function from which the NCF can be deduced \cite{landau}.  It was shown more recently \cite{premakumar} that the resulting equations for the NCF are equivalent to a classical electrodynamics problem: that of a point magnetic dipole source in the presence of the object.  More precisely, we place a fictitious dipole of strength $\vec{\mu}$ and frequency $\omega$ at the point $\vec{x}'$ and compute the fictitious field at $\vec{x}$.  This yields $\vec{B}_f(\vec{x},\vec{x}')$ where $\vec{x}$ is the observation point and $\vec{x}'$ is the source point.  Then the NCF at the frequency $\omega$ is given by  
\begin{equation}
\label{eq:ncf}
    \langle B_i(\vec{x}) B_j(\vec{x}')\rangle_\omega = \frac{\hbar}{\mu_j} \textrm{Im} \,  B_{ind,i}(\vec{x},\vec{x}') \coth({\hbar \omega / 2 k_B T}).
\end{equation} 

 $B_i(\vec{x})$ and $ B_j(\vec{x}')$ are physical noise fields.  Here $ \vec{B}_{ind}(\vec{x},\vec{x}')$ is the induced field, that is the total field $\vec{B}_f(\vec{x},\vec{x}')$ in the fictitious problem, minus the self field (the field in the absence of the object). $\vec{B}_{ind}(\vec{x},\vec{x}')$ is a kind of Green's function, but since we wish to stress the analogy to a magnetostatic problem, we prefer the present notation.  The coth function results from the bosonic character of the photons and it includes the emission of thermally excited photons from the metal that are absorbed by the qubit and the reverse process from the qubit (which is not necessarily in thermodynamic equilibrium) to the object. Both contribute to decoherence. The units of the NCF are erg-s/cm$^3$ in the CGS units used in this paper.  The equations to be solved for $\vec{B}_{f}(\vec{x},\vec{x}')$ are the standard Maxwell equations, supplemented by the boundary condition that the magnetic field is continuous at the surface of the object and by the constitutive relation $\vec{J} = \sigma \vec{E}$ for the current $\vec{J}$ as a function of the electric field $\vec{E}$ inside the object.  $\sigma$ is the conductivity.   

\subsection{Simplification of Maxwell Equations}
\label{subsec:simplification}
The equations to be solved for the fictitious fields in the frequency domain are
\begin{align}
\label{eq:maxwell}
    \nabla_{\vec{x}} \cdot \vec{B}_f &= 0 &\nabla_{\vec{x}} \times \vec{E}_f - i \frac{\omega}{c}\vec{B}_f &= 0 \\
    \nabla_{\vec{x}} \cdot \vec{E}_f &= 0  &\nabla_{\vec{x}} \times \vec{B}_f + i \frac{\omega}{c} \vec{E}_f &= \frac{4 \pi}{c}
\vec{J}.    
\end{align}
$\vec{E}_f$ is the fictitious electric field associated with $\vec{B}_f$. Outside the metal we have the dipole current $\vec{J}(\vec{x})=\nabla \delta^3(\vec{x}-\vec{x}') \times \vec{\mu}$ while inside the metal $\vec{J} = \sigma \vec{E}$.  All quantities have the time dependence $\exp(-i \omega t)$.   We will work in the frequency domain henceforth.

For the nanoscale qubit applications that are the subject here, there are some simplifications of these equations that can be obtained by looking at some characteristic length and time scales.  The frequencies of interest are at the operating frequency of the qubit or below, which gives the inequality $\omega < 10^{10}$ Hz.  This yields a lower bound on the vacuum wavelength: 
$\lambda = 2 \pi c/  \omega > 18.8 $cm.  A typical conductivity \cite{morello} would be in the range $\sigma = 1.6 \times 10^7 S/m = 1.44 \times 10^{17} s^{-1}$.  We will take this as a representative value for illustrative purposes below. These values yield the skin depth $\delta = c/ \sqrt{2 \pi \sigma \omega} = 3.14 \times 10^{-4}$cm (around a micron).  In nanoscale semiconductor qubit devices, we may take a maximum radius $a_m$ for our object that satisfies $a_m < 10^{-5} cm$.  Summarizing our considerations we have the inequalities 
\begin{equation}
\label{eq:inequality}
a_m < \delta < \lambda.
\end{equation}

Some other conditions are important for the validity of the theory presented in this paper.  We need $\omega << 1/\tau $, where $\tau$ is the relaxation time of the electrons, since otherwise one cannot neglect the frequency dependence of the conductivity. Similarly the mean free path of the electrons must be short compared with $a_m$, since otherwise the spatial dependence of the relation between current and electric field cannot be neglected.  Finally $|\vec{x}|$ and $|\vec{x}'|$ must be small compared with $\lambda$ for the quasistatic approximation to be valid.  All of these conditions are normally satisfied in the spin qubit systems of interest here.

Superconducting qubit circuit elements can be larger than the skin depth $\delta$.  Our theory does not work for this case. However, it can still serve as the basis for some simple approximate solutions that we will mention below. 

Rewriting Eqs. 2 and 3, the Maxwell equations inside the metal using the variables $\lambda$ and $\delta$, we have 
\begin{align}
    \nabla_{\vec{x}} \cdot \vec{B_f} &= 0 &\nabla_{\vec{x}} \times \vec{E}_f - \frac{2 \pi i}{\lambda}\vec{B}_f &= 0 \\
    \nabla_{\vec{x}} \cdot \vec{E}_f &= 0  &\nabla_{\vec{x}} \times \vec{B}_f + \frac{2 \pi i}{\lambda} \vec{E}_f &=  \frac{\lambda}{\pi \delta^2} \vec{E}_f.  
    \end{align}
Eliminating $\vec{E}_f$ from these equations gives
\begin{equation*}
 \nabla_{\vec{x}}^2 \vec{B}_f = (-\frac{2i}{\delta^2} -\frac{4 \pi^2}{\lambda^2}) \vec{B}_f.  
\end{equation*}
Using our inequalities to neglect terms of order $\delta^2/\lambda^{2}$ (the quasistatic approximation) we find
\begin{equation}
 \nabla_{\vec{x}}^2 \vec{B}_f = - \frac{2i}{\delta^2} \vec{B}_f
\end{equation}
in the metal.  Outside the object we have simply $\nabla_{\vec{x}}^2 \vec{B}_f =0$ and  $\nabla_{\vec{x}} \times \vec{B}_f= 4 \pi \nabla \delta^3(\vec{x}-\vec{x}') \times \vec{\mu}/c$.  

Thus far these are rather standard approximations.  Further progress may be made by solving the problem in two stages.  The solution $\vec{B}_d$ for the dipole problem in the absence of the sphere is  
\begin{equation}
\label{eq:dipole}
\vec{B}_d(\vec{x},\vec{x}') = \frac{3(\vec{x}-\vec{x}')(\vec{x}-\vec{x}')\cdot\vec{\mu}
-\vec{\mu}|\vec{x}-\vec{x}'|^2}{|\vec{x}-\vec{x}'|^5},
\end{equation}
and we write the total solution as the sum of the dipole field and the induced field:
\begin{equation}
\label{eq:expand}
\vec{B}_f(\vec{x},\vec{x}')= \vec{B}_d(\vec{x},\vec{x}') + \vec{B}_{ind}(\vec{x},\vec{x}'). 
\end{equation}
The equation for the total field is then  
\begin{equation}
\label{eq:it}
 (\nabla_{\vec{x}}^2 +\frac{2i}{\delta^2}) (\vec{B}_d + \vec{B}_{ind}) = 0.
\end{equation}
Expanding $\vec{B}_{ind}$ in powers of $\delta^{-2}$ we find that 
\begin{equation}
\label{eq:_simplified}
\nabla_{\vec{x}}^2 \vec{B}_{ind} = - \frac{2i}{\delta^2} \vec{B}_d,
\end{equation}
correct to order $a_m^2/\delta^2$.  Since $\vec{B}_d$ is given by Eq. \ref{eq:dipole}, this is simply a Poisson equation for the components of $\vec{B}_{ind}$, which is the field that enters the NCF.  Actually the result for the NCF is accurate to order $a_m^4/\delta^4$ since the fourth-order term in $\vec{B}_{ind}$ is real, and the NCF depends only on the imaginary part, as is seen from Eq. \ref{eq:ncf}. Since $\delta^{-2} = 2 \pi \sigma \omega / c^2$, the formula also immediately implies that the NCF is linearly proportional to the conductivity $\sigma$ and to the frequency $\omega$. 
\subsection{Solution for Arbitrary Shape}
\label{subsec:solution}
The fictitious dipole at $\vec{x}'$ sets up a vector potential 
\begin{equation}
\label{eq:ad}
    \vec{A}_d(\vec{x}'',\vec{x'}) = 
    \frac{\vec{\mu} \times (\vec{x}'' - \vec{x}')}{|\vec{x}'' - \vec{x}'|^3}+\nabla_{\vec{x}''}f_d(\vec{x}''), 
\end{equation}
where we have indicated the gauge ambiguity explicitly by including the function $f_d$.  Inside the object we have 
\begin{equation}
\label{eq:ed}
    \vec{E}_d(\vec{x}'',\vec{x'}) =
    \frac{i\omega}{c} \vec{A}_d(\vec{x}'') =
    \frac{1}{\sigma}\vec{J}(\vec{x}''),
\end{equation}
since we are using the temporal gauge.

$\nabla_{\vec{x}''} \cdot \vec{E}_d(\vec{x},\vec{x'})'' = 0$ so $\nabla_{\vec{x}''} \cdot \vec{A}_d(\vec{x},\vec{x'}'') =0$.
Furthermore, 
\begin{equation*}
   \nabla_{\vec{x}''}  \cdot \frac{\vec{\mu} \times (\vec{x}'' - \vec{x}')}{|\vec{x}'' - \vec{x}'|^3} =0,
\end{equation*}
so $\nabla_{\vec{x}''}^2 f_d(\vec{x}'',\vec{x'})=0$.  This means that $f_d$ satisfies a Laplace equation.  To determine the boundary condition, note that $\vec{J}(\vec{x}'') \cdot \hat{n} = 0$, where $\hat{n}$ is the outward-pointing normal vector to the object. Hence
\begin{equation*}
    \hat{n} \cdot \nabla_{\vec{x}''} f_d(\vec{x}'') = - \hat{n} \cdot \frac{\vec{\mu} \times (\vec{x}'' - \vec{x}')}{|\vec{x}'' - \vec{x}'|^3} 
    \end{equation*}
    
This is a Neumann boundary condition, so $f_d(\vec{x}'')$ is determined uniquely up to an unimportant global constant.   
From Eq.\ref{eq:ed} we have that
\begin{equation}
\label{eq:jd}
\vec{J}(\vec{x}'',\vec{x}') = \frac{i \omega \sigma}{c} \left[ \frac{\vec{\mu} \times (\vec{x}'' - \vec{x}')}{|\vec{x}'' - \vec{x}'|^3} + \nabla_{\vec{x}''} f_d(\vec{x}'',\vec{x}') \right]
\end{equation}
This current creates the induced field $\vec{B}_{ind}(\vec{x},\vec{x}')$.  Note however, that the second term in $\vec{J}$ is purely longitudinal, and for a source that occupies a \textit{finite} region, this part of the current does not contribute to the induced field.  At this point we may apply the Biot-Savart law together with Eq.\ref{eq:jd} and obtain
\begin{equation}
\label{eq:bind1}
\vec{B}_{ind} (\vec{x},\vec{x}') = \frac{i}{ 2 \pi \delta ^2} \int d^3 x'' \frac{\vec{\mu} \times (\vec{x}- \vec{x}') }
{ |(\vec{x}- \vec{x}')|^3} \times \frac{(\vec{x}- \vec{x}'') }
{ |(\vec{x}- \vec{x}'')|^3}
\end{equation}
Here the limits of the integral run only over the volume of the metal object. However, it is obviously equally valid if there are multiple objects, their effects being additive.  Note that $\vec{x}'$ is a parameter in this equation. Of course $\vec{x}'$ plays a role of equal importance to $\vec{x}$ once it is substituted back into Eq. \ref{eq:ncf}.

Eq.\ref{eq:bind1} is a remarkably simple expression.  It gives an explicit method for the calculation of $\vec{B}_{ind}$, which when substituted into Eq.\ref{eq:ncf} gives directly the NCF.  It can therefore serve as the basis for a straightforward numerical calculation of the NCF for a finite object of arbitrary shape.  For most practical purposes, one needs to compute only a few multipole moments of the three integrals to obtain a serviceable answer. 

At this point it is also useful to ask what happens when the assumption $\delta >> a_{m}$ breaks down.  We do not have exact results, but we can say a few things based on the above.  If $\delta \gtrsim a_{m}$ then the expansion of the NCF in powers of $a_m^2/\delta^2 $ can be continued by repeated substitution into Eq. \ref{eq:it}, though the resulting differential equation for the next-order term is complicated.  In the opposite limit when $\delta << a_{min}$, where $a_{min}$ is the minimum radius of the object, then the field does not penetrate into the interior of the object but only to a distance $\delta$.  We may apply Eq. \ref{eq:bind1} directly but replace the volume of integration by a shell of width $\delta$ from the surface to obtain an approximate result.

\onecolumngrid

\section{Multipole Expansion for the Sphere}
\label{sec:sphere}
In this section, we derive the multipole expansion for the NCF of a spherical conductor of radius $a$, taking advantage of the symmetry of the problem.  We introduce the notation $\vec{x}=(x_1,x_2,x_3) = (r,\theta,\phi)$ in Cartesian and spherical coordinates and $\partial_i=\partial/\partial x_i$ and similarly for $x_i'$ and $x_i''$.  

\subsection{Induced Magnetic Field}
\label{subsec:balt}

Outside the sphere there are no currents that affect $\vec{B}_{ind}$ so we may define a scalar magnetic potential $\psi_{ind}(\vec{x})$ that determines $\vec{B}_{ind}$ through  $\vec{B}_{ind}(\vec{x}) = - \nabla \psi_{ind}(\vec{x})$.  Hence
\begin{equation*}
    \vec{x} \cdot \vec{B}_{ind} = - r \frac{\partial \psi_{ind}}{\partial r}. 
\end{equation*}
and we have a Poisson equation for the radial part of the field 
\begin{equation*}
    \nabla_{\vec{x}}^2(\vec{x} \cdot \vec{B}_{ind} (\vec{x})) =
    - \frac{4 \pi}{c} \vec{x} \cdot \nabla \times \vec{J} (\vec{x})
\end{equation*}
with the solution
\begin{equation*}
    \vec{x} \cdot \vec{B}_{ind}(\vec{x}) = 
    \frac{1}{c} \int_{r'' \leq a} \frac{d^3 x''}{|\vec{x} - \vec{x}''|}
    \vec{x}'' \cdot \left(\nabla'' \times \vec{J}(\vec{x}'')\right) = 
    -r \frac{\partial \psi_{ind}(\vec{x})}{\partial r}. 
\end{equation*}

So we may write the induced field as 
\begin{equation*}
    \vec{B}_{ind} (\vec{x},\vec{x}') =
    \frac{1}{c} \nabla_{\vec{x}}
    \left[\int^r \frac{dr}{r}
    \int_{object} \frac{d^3 x''}{|\vec{x} - \vec{x}''|}
    \vec{x}'' \cdot \nabla'' \times \vec{J}(\vec{x}'') \right].
\end{equation*}
$\vec{J}$ depends parametrically on $\vec{x}'$.  Referring to Eqs. \ref{eq:ad} and \ref{eq:ed} we see that $f_d$ does not contribute to the field.  This is an application of the principle that a longitudinal current does not produce a magnetic field.  Finally we have the formula
\begin{equation}
\label{eq:bind}
    \vec{B}_{ind} (\vec{x},\vec{x}') = \frac{i}{2 \pi \delta^2}
    \nabla
    \left[\int^r \frac{dr}{r} 
    \int_{object} \frac{d^3 x''}{|\vec{x} - \vec{x}''|}
    \vec{x}'' \cdot \vec{B}_d(\vec{x}'')\right].  
\end{equation}
Keeping in mind the relations
\begin{equation}
\label{eq:partial}
\partial_i \frac{1}{|\vec{x} - \vec{x}'|}
= - \partial_i' \frac{1}{|\vec{x} - \vec{x}'|} 
= - \frac{x_i-x_i'}{|\vec{x} - \vec{x}'|^3},
\end{equation}
and using Eq. \ref{eq:partial} twice to rewrite Eq. \ref{eq:dipole} as
\begin{equation}
\label{eq:dipolea}
\vec{B}_d(\vec{x}'',\vec{x}') = - \partial_i'' \mu_j \partial_j'
\frac{1}{|\vec{x}''-\vec{x}'|},
\end{equation}
we find the component version of Eq. \ref{eq:bind} to be
\begin{equation}
\label{eq:bindk}
B_{ind,k} (\vec{x},\vec{x}') = - \frac{i}{2 \pi \delta^2}
    \mu_j \partial_j' \partial_k
    \left[\int_r \frac{1}{r} dr 
    \int_{object} d^3 x'' x_i''  \frac{1}{|\vec{x} - \vec{x}''|} \partial_i''  \frac{1}{|\vec{x}'' - \vec{x}'|} \right].
\end{equation}
Summation over repeated Cartesian indices is understood.
Now we use the standard relation
\begin{equation}
\label{eq:ylm}
    \frac{1}{|{\vec{x}-\vec{x^\prime}}|}=\sum^\infty_{\ell = 0}\sum^{\ell}_{m=-\ell}\frac{4\pi}{2\ell+1}\frac{r^{\prime\ell}}{r^{\ell+1}}
    Y^*_{lm}(\theta',\phi')Y_{lm}(\theta,\phi)
\end{equation}
valid for $r'<r$. 
Substituting Eq. \ref{eq:ylm} into Eq. \ref{eq:bindk} we find
\begin{align}
    \nonumber
    B_{ind,k}(\vec{x},\vec{x}')={}&-\frac{i}{2\pi\delta^2}
    \mu_j\partial^\prime_j\partial_k \bigg[ \int_0^r r^{-1}dr \int d^3x^{\prime\prime}x_i^{\prime\prime} \times
    \sum_{\ell=0}^{\infty}\sum_{m=-\ell}^{\ell} \frac{4\pi}{2l+1}\frac{
    (r^{\prime\prime})^l}{r^{l+1}}
    Y^*_{\ell m}(\theta^{\prime\prime},\phi^{\prime\prime})
    Y_{\ell m}(\theta,\phi)  \times \\
    \nonumber
    {}&\sum_{\ell^\prime=0}^{\infty}
    \sum_{m^\prime=-\ell^\prime}^{l^\prime} \frac{4\pi}{2\ell^\prime+1}\partial_i^{\prime\prime}
    \frac{(r^{\prime\prime})^{\ell^\prime}}{(r^\prime)^{l^\prime+1}}Y^*_{\ell^\prime m^\prime}(\theta^{\prime},\phi^{\prime})Y_{\ell^\prime m^\prime}(\theta^{\prime\prime},\phi^{\prime\prime})\bigg]\\
    \nonumber
    ={}&\frac{i}{2\pi\delta^2}\mu_j \sum_{\ell,m} \frac{4\pi}{(l+1)(2l+1)}\partial_k
    \left[\frac{Y_{\ell m}(\theta,\phi)}{r^{\ell+1}}\right] \times
    \sum_{\ell^\prime,m^\prime}\frac{4\pi}{(2l^\prime+1)}
    \partial_j^\prime\left[\frac{Y^*_{\ell^\prime m^\prime}}{(r^\prime)^{\ell^\prime+1}}\right] \times \\
    \label{eq:comp}
    {}&\int d^3x^{\prime\prime}x_i^{\prime\prime}(r^{\prime\prime})^\ell Y^*_{\ell m}(\theta^{\prime\prime},\phi^{\prime\prime})\partial_i^{\prime\prime}\left[(r^{\prime\prime})^{\ell^\prime}Y_{\ell^\prime m^\prime}(\theta^{\prime\prime},\phi^{\prime\prime})\right]
\end{align}
This rather complicated-looking formula will serve as the basis for the multipole expansion.
\subsection{Definitions and Auxiliary Quantities}
The vector spherical harmonics as defined by Barrera \textit{et al.} \cite{barrera} are
\begin{equation}
\label{eq:vec}
    \vec{Y}_{\ell m} (\vec{x}) = \hat{r} Y_{\ell m} (\theta,\phi)
\end{equation}
and
\begin{equation}
\vec{\Psi} (\vec{x}) = r \nabla Y_{\ell m} (\theta,\phi).
\end{equation}
We will need the fact that
\begin{equation}
\label{eq:yder}
\partial_k [r^{\ell} Y_{\ell m } (\theta, \phi)] =
\ell r^{\ell - 1} \vec{Y}_{\ell m, k} (\vec{x}) + r^{\ell -1}
\vec{\Psi}_{\ell m , k} (\vec{x}).
\end{equation}
We also make the new definitions
\begin{align}
\label{eq:slm}
    \vec{S}_{\ell m} (\vec{x}) 
    & = 
    (\ell+1)  \vec{Y}_{\ell m} (\theta,\phi) -  \vec{\Psi}_{\ell m} (\theta,\phi) \\ 
    &=
    (\ell + 1)  \hat{r} Y_{\ell m} (\vec{x}) -
    \hat{\theta} \frac{\partial}{\partial \theta}  Y_{\ell m} (\theta,\phi) - \hat{\phi} 
    \frac{1}{\sin \theta}\frac{\partial}{\partial\phi}Y_{\ell m}(\theta,\phi) 
\end{align}
and 
\begin{equation}
\label{eq:aell}
 A_{\ell}=\frac{\ell}{(\ell+1)(2\ell+1)^2(2\ell+3)}.
\end{equation}

\subsection{Collapse to a Multipole Expansion}
Eq. \ref{eq:yder} now allows us to express the derivatives in Eq. \ref{eq:comp} as vector spherical harmonics, while Eq. \ref{eq:vec} shows how to re-express them as scalar harmonics for the variable $\vec{x}''$ and finally apply the orthogonality property of the $Y_{\ell m}$ to perform the integral as follows.  Performing the differentiations in Eq.\ref{eq:comp} we find:
\begin{align*}
    \vec{B}_{ind}(\vec{x},\vec{x}')={}&\frac{i}{2\pi\delta^2}
    \sum_{l,m} \frac{4\pi}{(l+1)(2l+1)}
    \left[-\frac{l+1}{r^{l+2}}\vec{Y}_{lm}(\vec{x})+\frac{1}{r^{l+2}}\vec{\Psi}_{lm}(\vec{x})\right] \times \\
    {}&\sum_{l',m'}\frac{4\pi}{(2l^\prime+1)}\vec{\mu}\cdot\left[-\frac{l^\prime+1}{(r^\prime)^{l^\prime+2}}\vec{Y^*}_{l^\prime m^\prime}(\vec{x^\prime})+\frac{1}{(r^\prime)^{l^\prime+2}}\vec{\Psi^*}_{l^\prime m^\prime}(\vec{x^\prime})\right] \times \\
    {}&\int d^3x^{\prime\prime}(r^{\prime\prime})^lY^*_{lm}(\theta^{\prime\prime},\phi^{\prime\prime})
    \vec{x^{\prime\prime}}\cdot \left[l^\prime(r^{\prime\prime})^{l^\prime-1}\vec{Y}_{l^\prime m^\prime}(\vec{x^{\prime\prime}})+(r^{\prime\prime})^{l^\prime-1}\vec{\Psi}_{l^\prime m^\prime}(\vec{x^{\prime\prime}})\right].
\end{align*}
    
The integral can now be evaluated:
\begin{align*}
    &\int d^3x^{\prime\prime}(r^{\prime\prime})^l Y^*_{lm}(\theta^{\prime\prime},\phi^{\prime\prime})\vec{x^{\prime\prime}}\cdot \left[l^\prime(r^{\prime\prime})^{l^\prime-1}\vec{Y}_{l^\prime m^\prime}(\vec{x^{\prime\prime}})+(r^{\prime\prime})^{l^\prime-1}\vec{\Psi}_{l^\prime m^\prime}(\vec{x^{\prime\prime}})\right]\\
    &=\int d^3x^{\prime\prime}Y^*_{lm}(\theta^{\prime\prime},\phi^{\prime\prime})\vec{x^{\prime\prime}}\cdot\left[l^\prime r^{{\prime\prime}^{(l^\prime+l-1)}}Y_{l^\prime m^\prime}(\theta^{\prime\prime},\phi^{\prime\prime})\hat{x^{\prime\prime}}\right]\\
    &=l^\prime \int_0^a dr^{\prime\prime} r^{{\prime\prime}^{(l^\prime+l-1)}}r^{\prime\prime} \int d\Omega^{\prime\prime}Y^*_{lm}(\Omega^{\prime\prime})Y_{lm}(\Omega^{\prime\prime})\\
    &=\frac{l^\prime a^{l+l^\prime+3}}{l+l^\prime+3} \delta_{l,l^\prime}\delta_{m,m^\prime},
\end{align*}
where $a$ is the radius of the sphere.  The key technical issue in the derivation is to move the derivatives entirely to the $\vec{x}$ and $\vec{x}'$ variables so that the orthogonality relation can be used to perform the integration.  Again, this result is only valid when $a<<\delta$.  However, it can be used to obtain an approximate answer for the case $\delta<<a$; we can change the lower limit on the angular integral to $a-\delta$ which leads to the replacement of $a^{\ell + \ell ' +3} $ by  $a^{\ell + \ell ' +3} - (a-\delta) ^{\ell + \ell ' +3}$ in the last line.  

Simplifying somewhat further, we obtain
\begin{align*}
    \vec{B}_{ind}(\vec{x},\vec{x}')={}&\frac{i}{2\pi\delta^2} \sum_{l=0}^{\infty}\sum_{m=-\ell}^{\ell} \frac{16\pi^2}{(l+1)(2l+1)^2} \left(\frac{l a^{2\ell+3}}{2\ell+3}\right)\left[-\frac{\ell+1}{r^{\ell+2}}\vec{Y}_{\ell m}(\vec{x})+\frac{1}{r^{l+2}}\vec{\Psi}_{\ell m}(\vec{x})\right]\\
    &\vec{\mu}\cdot\left[-\frac{\ell ^\prime+1}{(r^\prime)^{\ell^\prime+2}}\vec{Y^*}_{\ell^\prime m^\prime}(\vec{x})+\frac{1}{(r^\prime)^{\ell^\prime+2}}\vec{\Psi}^*_{\ell^\prime m^\prime}(\vec{x})\right]\\
    ={}&\frac{8\pi}{\delta^2}\sum_{\ell=0}^{\infty}\sum_{m=-\ell}^{\ell} \frac{l}{(\ell+1)(2\ell+1)^2(2\ell+3)}\frac{a^{2\ell+3}}{(rr^\prime)^{\ell+2}}\left[(\ell +1)\vec{Y}_{\ell m}(\theta,\phi)-\vec{\Psi}_{\ell m}(\theta,\phi)\right]\nonumber\\
    {}&\vec{\mu}\cdot\left[(\ell+1)\vec{Y^*}_{\ell m}(\theta^\prime,\phi^\prime)-\vec{\Psi}^*_{\ell m}(\theta^\prime,\phi^\prime)\right].
\end{align*}
 This may be put into a compact form
    \begin{equation}
    \label{eq:compact}
       \vec{B}_{ind}(\vec{x},\vec{x}')=\frac{8i\pi}{\delta^2}\sum_{\ell=1}^{\infty}\sum_{m=-\ell}^{m=\ell}A_l\frac{a^{2\ell+3}}{(rr')^{\ell+2}} \vec{S}_{\ell m}(\theta,\phi) \vec{\mu}\cdot\vec{S^*}_{\ell m}(\theta',\phi'), 
    \end{equation}
 using the definitions in Eqs. \ref{eq:slm} and \ref{eq:aell}. This is the second new result in this paper.  As it stands, it is a closed form solution for a problem in classical electromagnetic theory.  Then substitution into Eq. \ref{eq:ncf} yields immediately the magnetic NCF.  Since $A_\ell=0$ when $\ell=0$, we see that the expansion begins at $\ell=1$.  This of course is just the non-existence of the monopole moment.  It shows explicitly that the asymptotic long-distance behavior of the NCF is $r^{-3} (r')^{-3}$, as is to be expected once the dipole analogy for the problem is accepted.  Once again we note that the theory only holds for $r,r' << \lambda$ and asymptotics must be applied only with this proviso.
 
 \subsection{Dimensionless Form}
 \label{subsec:dimensionless}
 Substituting Eq.\ref{eq:compact} into Eq.\ref{eq:ncf} we find
 \begin{equation}
 \label{eq:bb}
    \langle B_i(\vec{x}) B_j(\vec{x}')\rangle_\omega 
    =
    \frac{8\pi\hbar}{\delta^2 a \mu_j}
    \coth\bigg(\frac{\hbar \omega}{2 k_B T}\bigg)
    \sum_{\ell=1}^{\infty} \sum_{m=-\ell}^{m=-\ell}
    A_{\ell} \, \bigg(\frac{r}{a}\bigg)^{-\ell -2} \, 
    \bigg(\frac{r'}{a}\bigg)^{-\ell -2}\,
    S_{\ell m,i}(\theta,\phi) \,  
    S^*_{\ell m,j}(\theta',\phi'),  
 \end{equation}
and we note that the Cartesian components $S_{\ell m , i}$ are dimensionless functions of angle only. The coth function is very important for identifying EWJN experimentally, but here we are mainly interested in the spatial dependence so we set $T=0$ which gives $\coth(\hbar \omega / 2 k_B T) =1$. Then we have the prefactor $8\pi\hbar/\delta^2 a$ which for our illustrative values $\sigma =1.44 \times 10^{17}/$s, $\omega = 10^{10}/$s and $a=10^{-5}$cm is $8\pi\hbar/\delta^2 a = 2.65 \times 10^{-14}$ erg-s/cm$^3$.  To get some idea of the physical meaning of this number, an electron spin qubit in a noise field of this magnitude would have a relaxation rate of about $1/T_1 \approx 2$s$^{-1}$, which is a typical value for spin qubits in silicon nanodevices. 

We are thus motivated to define the dimensionless functions $F_{ij}$ by
\begin{equation}
    \label{eq:dimensionless}
    F_{ij}(\vec{x},\vec{x}') = 
    \sum_{\ell=1}^{\infty} \sum_{m=-\ell}^{m=-\ell}
    A_{\ell} \, \bigg(\frac{r}{a}\bigg)^{-\ell -2} \, 
    \bigg(\frac{r'}{a}\bigg)^{-\ell -2}\,
    S_{\ell m,i}(\theta,\phi) \,  
    S^*_{\ell m,j}(\theta',\phi'), 
\end{equation}
which will be investigated numerically below.

\subsection{Miniaturization}
Eq.\ref{eq:bb} is the most convenient for understanding the scaling of the NCR for miniaturization.  We may generally assume a feature size $D$ such that $ D \approx r \approx a$ in a device. When $D$ is around 100 nm, then any spin coherence time $\tau$ is of order 1 s and what Eq.\ref{eq:bb} shows is that $\tau \propto 1/D$.  Thus smaller devices correspond to shorter coherence times and the times are inversely proportional to the inverse first power of the feature size.        

\subsection{Transformation Properties}
\label{subsec:tensor}
The underlying problem has spherical symmetry.  The NCF is a second-rank tensor so this symmetry puts strong restrictions on its form.  Let us define a rotation $\bm{R}$ with matrix representation $R_{ij}$ drawn from $SO(3)$. $\bm{R}$ is parameterized by the usual Euler angles.  We use active transformations $\bm{R}$ because we do not want to introduce a new coordinate system. Then the NCF satisfies
\begin{equation}
\label{eq:rotation}
  F_{ij}(\bm{R}\vec{x},\bm{R}\vec{x}')
  =    
  \sum_{i,j} R_{ik} R_{jl} 
  F_{kl}(\vec{x},\vec{x}'), 
\end{equation} 
and we will see illustrations of this relation below.

\vspace{0.1 in.}

\twocolumngrid
\section{Results for the Local Noise Correlation Function of a Metallic Sphere}
\label{sec:lncf}
\subsection{General Formula}
\label{subsec:general_lncf}
We proceed to find the formulas for the dimensionless form of the local NCFs $\langle B_z(\vec{x}) B_z(\vec{x}) \rangle_{\omega}$ and $\langle B_x(\vec{x}) B_x(\vec{x}) \rangle_{\omega}$. These are special cases of Eq. \ref{eq:dimensionless} when $\vec{x^\prime}=\vec{x}$ and $i=j=z$ or $i=j=x$. Because we want to illustrate how quickly the multipole moments converge, we will sum over $\ell$ only up to $L$.  Thus we arrive at
\begin{equation}
    \label{eqn:local_zz}
    F^{(L)}_{zz} (\vec{x},\vec{x}) = \sum_{\ell=1}^{L}\sum_{m=-\ell}^{m=\ell} A_\ell \left(\frac{a}{r}\right)^{2\ell+4}
    \hat{z}\cdot\vec{S}_{\ell m}(\theta,\phi) \,  \hat{z}\cdot\vec{S^*}_{\ell m}(\theta,\phi)
\end{equation}
and
\begin{equation}
    \label{eqn:local_xx}
    F^{(L)}_{xx} (\vec{x},\vec{x}) = \sum_{\ell=1}^{L}\sum_{m=-\ell}^{m=\ell}
    A_\ell \left(\frac{a}{r}\right)^{2\ell+4}
    \hat{x}\cdot\vec{S}_{\ell m}(\theta,\phi) \, \hat{x}\cdot\vec{S^*}_{\ell m}(\theta,\phi)
\end{equation}

We will be plotting the equations on the $x-z$ plane at $y=0$, which means $\phi=0$. We will vary $L$ from $L=1$ to $L=5$ to understand the contribution of the first few multipoles to the noise correlation function. 

\subsection{Angular Patterns for the local NCF}

To get some physical insight into the formulas for the NCF, we now compute numerically and then plot some components of the local tensor.   

Substituting Eq.\ref{eq:slm} into Eq.\ref{eqn:local_zz} returns
\begin{align}
    F^{(L)}_{zz}(\vec{x},\vec{x}) = {}& \sum_{l=\ell}^{L}\sum_{m=-\ell}^{m=\ell}A_\ell\bigg(\frac{a}{r}\bigg)^{2\ell +4}\nonumber\\
    {}&\bigg[(\ell +1) \cos\theta \, Y_{\ell m}(\theta,0)+\sin\theta\frac{\partial}{\partial\theta}Y_{\ell m}(\theta,0)\bigg]\nonumber\\
    {}&\bigg[(\ell +1) \cos\theta \, Y^*_{\ell m}(\theta,0)+\sin\theta\frac{\partial}{\partial\theta}Y^*_{\ell m}(\theta,0)\bigg]
\end{align}

This formula is expressed entirely in terms of tabulated functions.  

\begin{figure}[hbt!]
    \centering
    \includegraphics[width=\columnwidth]{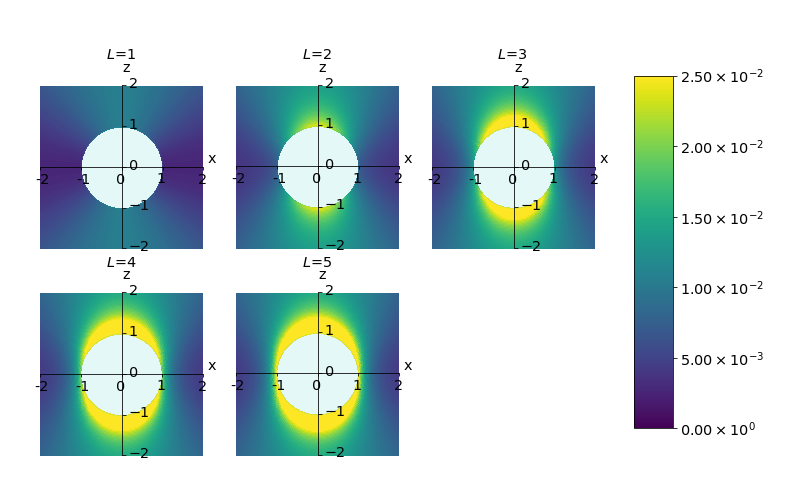}
    \caption{Dimensionless local noise correlation function $F^{(L)}_{zz}(\vec{x},\vec{x})\times (r/a)^6$.  The unit of distance is $a$, the radius of the sphere.  Shown is the function obtained for $L=1$ to $L=5$, where $L$ is the number of multipoles included in the sum. Deep-blue color indicates regions of very small positive correlation and yellow color shows region of high positive correlation.}
    \label{fig:local_bzbz_0}
\end{figure}

For better visualization of the angular dependence of the local NCF, we will remove the lowest-order dependence on distance by multiplying it by $(r/a)^6$. Fig. \ref{fig:local_bzbz_0} shows 
$F^{(L)}_{zz}(\vec{x},\vec{x}) \times (r/a)^6 $.

The angular pattern for the local NCF shown in Fig.\ref{fig:local_bzbz_0} can be understood using the analogy to the classical problem \cite{premakumar}.  A dipole that points in the z-direction that oscillates at the frequency $\omega$ is placed at $\vec{x}$.  It induces currents in the object, which in turn produce the induced field component $B_{ind,z}$.  Up to constant, this is what is shown in Fig.\ref{fig:local_bzbz_0}.  

For example, for Fig.\ref{fig:local_bzbz_0}(a), let $\vec{x}=(0,0,d)$ with $d>>a$ so that the pure dipole approximation ($L=1$) is valid for the induced field of the object.  At the object, the  field of the original dipole is in the +z-direction and it is strong, since the object lies along the direction of the dipole. The induced dipole is also in the $z$-direction and it is strong since it is proportional to the applied field.  $B_{ind,z}(\vec{x})$ is therefore large and positive. Now let  $\vec{x}=(d,0,0)$ with $d>>a$.  The  field of the original dipole is slightly less strong at the object, since the object lies along the direction perpendicular to the dipole, so it is in the return field. The induced dipole is in the $-z$-direction and it is weaker.  Hence we find a smaller result at  $\vec{x}=(d,0,0)$ than at $\vec{x}=(0,0,d)$, which accounts for the anisotropy in the results.  Going beyond the dipole approximation when $d\geq a$, so that $L>1$, we must take into account that the currents in the object are strong in the parts of the sphere that are near the dipole, and weaker as we move farther away.  This amplifies the original mechanism since when $d$ is not much greater than $a$, there is considerable cancellation of the $x$-component of the return field but not when it is on the $z$-axis.  This accounts for the increasing anisotropy as more multipole moments are included in the calculation.                         
To illustrate the workings of the rotational symmetry expressed by Eq.\ref{eq:rotation}, we will look at the $xx$ component of the local NCF.

Solving Eq.\ref{eqn:local_xx}, we get
\begin{align}
    F^{(L)}_{xx}(\vec{x},\vec{x}) = {}& \sum_{l=\ell}^{L}\sum_{m=-\ell}^{m=\ell}A_\ell\bigg(\frac{a}{r}\bigg)^{2\ell +4}\nonumber\\
    {}&\bigg[\sin\theta(\ell +1)Y_{\ell m}(\theta,0)-\cos\theta\frac{\partial}{\partial\theta}Y_{\ell m}(\theta,0)\bigg]\nonumber\\
    {}&\bigg[\sin\theta(\ell +1)Y^*_{\ell m}(\theta,0)-\cos\theta\frac{\partial}{\partial\theta}Y^*_{\ell m}(\theta,0)\bigg].
\end{align}

\begin{figure}[hbt!]
    \centering
    \includegraphics[width=\columnwidth]{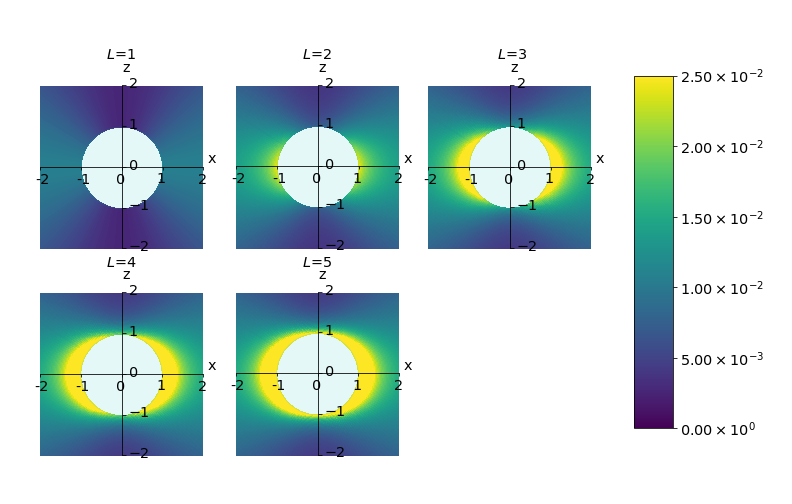}
    \caption{Dimensionless local noise correlation function $F^{(L)}_{xx}(\vec{x},\vec{x})\times (r/a)^6$.  The unit of distance is $a$, the radius of the sphere.  Shown is the function obtained for $L=1$ to $L=5$, where $L$ is the number of multipoles included in the sum. Deep-blue color indicates regions of very small positive correlation and yellow color shows region of high positive correlation.}
    \label{fig:local_bxbx_0}
\end{figure}

Fig.\ref{fig:local_bxbx_0} shows the$F^{(L)}_{xx}(\vec{x},\vec{x}) \times (r/a)^6 $ noise correlation.

Fig.\ref{fig:local_bxbx_0} can be understood simply by using Eq.\ref{eq:rotation} with $\bm{R}$ representing a rotation by angle $\gamma$ about the y-axis, so that $R_{xx}=R_{zz} = \cos\gamma$, $R_{xz}=-R_{zx} = \sin\gamma$, $R_{yy}= 1$ and all other components equal to 0.  Substitution of this form of $\bm{R}$ with $i=j=x$ into Eq.\ref{eq:rotation} leads immediately to 
\begin{equation*}
    \bm{R} (x,y,z) = ( - x\cos\gamma - x \sin\gamma,y,z\cos\gamma + x \sin\gamma)
\end{equation*}
and
\begin{align*}
    F_{xx}(\bm{R} (x,y,z), \bm{R} (x,y,z) ) 
    = & \nonumber \\
    \cos^2\gamma F_{xx}((x,y,z),(x,y,z)) + &
     \sin^2\gamma F_{zz}((x,y,z),(x,y,z)).
\end{align*}
In particular, when $\gamma = \pi /2$, we find \qquad \qquad \qquad \qquad \qquad
\linebreak
$ F_{xx}(\bm{R}\vec{x}, \bm{R} \vec{x}) =  F_{zz}(\vec{x},\vec{x})$.  This accounts for the evident relation 
between Figs.\ref{fig:local_bzbz_0} and \ref{fig:local_bxbx_0}, rotated by $\pi/2$ relative to each other.

\subsection{Implications for Quantum Computing}
\label{subsec:local}

The metal object may be a device element or (less likely) an accidental inclusion in an array of qubits.  Magnetic noise is important in many quantum computing platforms, but we will confine ourselves here to electron spin qubits.  Our aim is to show how our visualization of noise can help to develop recommendations for designers of quantum computing hardware and software.  In this subsection we focus on a hardware issue. 

We picture a single metallic object at the origin with qubits at position $\vec{x}$ in the vicinity. The local NCF creates decoherence only on local quantum amplitudes: single qubit $T_1$ and $T_2$.  Consider a qubit at $\vec{x}=(0,0,d)$.  The local NCF is a diagonal tensor at this point in the $(x,y,z)$ basis.  If the steady applied field $\vec{B}_{app}$ is in the $z$-direction, then $T_2$ is determined only by the $zz$ entry of the NCF tensor while $T_1$ is determined by the sum of the $xx$ and $yy$ entries. (Roughly speaking, we must multiply the NCF entries by $\mu_B^2/\hbar^2$ to get the decoherence times.) If the applied field is in the $\hat{n}$-direction, then $T_2$ and $T_1$ are respectively determined by $\sum_{ij} n_i n_j F_{ij}(\vec{x})$ and $\sum_{ij} [m_{1,i} m_{1,j}+m_{2,i} m_{2,j}] F_{ij}(\vec{x})$, where $\hat{m}_1$ and $\hat{m}_2$ are unit vectors orthogonal to each other and to $\hat{n}$.  At other qubit positions, we first apply Eq.\ref{eq:rotation} and then follow the same logic to obtain $T_1$ and $T_2$.

To give a simple example, let us say that we wish to maximize $T_2$, which is often the case, and the qubit is at $(d,0,0)$ with $d\gtrsim a$. Then we consult Figs.\ref{fig:local_bzbz_0} and \ref{fig:local_bxbx_0} and we see that the $zz$ entry of the local NCF is less than the $xx$ entry.  We conclude that we should use an applied field in the $z$-direction.

\section{Results for Nonlocal Noise Correlation Function}
\label{sec:nncf}
\subsection{General Formulas}
In this section, we investigate the nonlocal dimensionless NCF, that is,  $F^{(L)}_{zz}(\vec{x},\vec{x}')$ when  $\vec{x}\neq\vec{x}'$.  Using the classical analogy, we fix the position of the fictitious dipole at $\vec{x}'$  and compute $\vec{B}_{ind}(\vec{x},\vec{x}')$ at the observation point $\vec{x}$.  For purposes of illustration we will take $\vec{x}'=(0,0,|\vec{x}'|)$ at various distances from the origin. Again, the fictitious dipole sets up currents in the sphere which in turn create the induced field.  

Proceeding as in Sec. \ref{sec:lncf}, we have the following formulas for the nonlocal correlation functions:
\begin{align}
    F^{(L)}_{zz}(\vec{x},\vec{x}') = {}& \sum_{\ell=1}^{L} \sum_{m=-\ell}^{m=-\ell}
    A_{\ell}\bigg(\frac{a}{r}\bigg)^{\ell +2}\bigg(\frac{a}{r'}\bigg)^{\ell +2}\nonumber\\
    \label{eqn:nonlocal_zz}
    {}&\hat{z}\cdot\vec{S}_{\ell m}(\theta,\phi) \hat{z}\cdot\vec{S^*}_{\ell m}(\theta',\phi')
\end{align}
\begin{align}
    F^{(L)}_{xx}(\vec{x},\vec{x}') ={}& \sum_{\ell=1}^{L} \sum_{m=-\ell}^{m=-\ell}
    A_{\ell}\bigg(\frac{a}{r}\bigg)^{\ell +2}\bigg(\frac{a}{r'}\bigg)^{\ell +2}\nonumber\\
    \label{eqn:nonlocal_xx}
    {}&\hat{x}\cdot\vec{S}_{\ell m}(\theta,\phi) \hat{x}\cdot\vec{S^*}_{\ell m}(\theta',\phi')
\end{align}
\begin{align}
    F^{(L)}_{xz}(\vec{x},\vec{x}') ={}& \sum_{\ell=1}^{L} \sum_{m=-\ell}^{m=-\ell}
    A_{\ell}\bigg(\frac{a}{r}\bigg)^{\ell +2}\bigg(\frac{a}{r'}\bigg)^{\ell +2}\nonumber\\
    \label{eqn:nonlocal_xz}
    {}&\hat{x}\cdot\vec{S}_{\ell m}(\theta,\phi) \hat{z}\cdot\vec{S^*}_{\ell m}(\theta',\phi')
\end{align}

We choose the values $\vec{x^\prime}=(0,0,d)$, where $d=2a$ and $d=5a$. The values of $d$ are chosen to show how the various terms in the multipole expansion affect the nonlocal NCF.  

\subsection{$F^{(L)}_{zz}(\vec{x},\vec{x}')$ for various limits L on the sum}
\label{subsec:nncf_zz}
\begin{figure}[hbt!]
    \centering
    \includegraphics[width=\columnwidth]{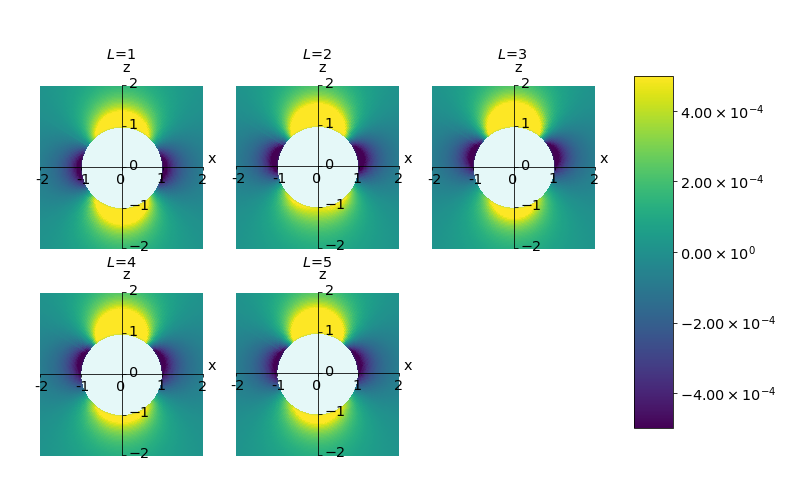}
    \caption{Dimensionless nonlocal noise correlation function $F^{(L)}_{zz}(\vec{x},\vec{x}')$.  Here $\vec{x}'=(0,0,2a)$ (at top center of the plots) is fixed and the plot is as a function of $\vec{x}$.  The unit of distance is $a$, the radius of the sphere.  Shown is the function obtained for $L=1$ to $L=5$, where $L$ is the number of multipoles included in the sum. Deep-blue color indicates regions of negative correlation and yellow color shows regions of positive correlation.}
    \label{fig:nonlocal_bzbz_m=2_0}
\end{figure}

\begin{figure}[hbt!]
    \centering
    \includegraphics[width=\columnwidth]{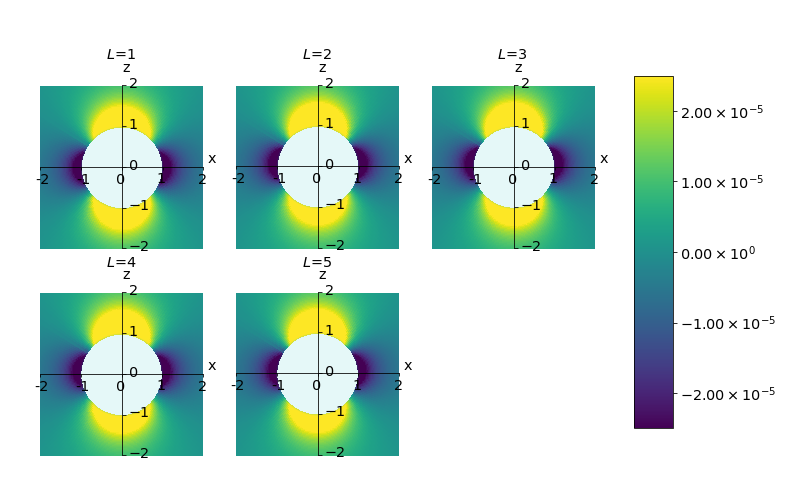}
    \caption{Dimensionless nonlocal noise correlation function $F^{(L)}_{zz}(\vec{x},\vec{x}')$.  Here $\vec{x}'=(0,0,5a)$ (off the plots) is fixed and the plot is as a function of $\vec{x}$.  The unit of distance is $a$, the radius of the sphere.  Shown is the function obtained for $L=1$ to $L=5$, where $L$ is the number of multipoles included in the sum. Deep-blue color indicates regions of negative correlation and yellow color shows regions of positive correlation.}
    \label{fig:nonlocal_bzbz_m=5_0}
\end{figure}

Expressing Eq.\ref{eqn:nonlocal_zz} in terms of the usual spherical harmonics, we get
\begin{align}
    \label{eqn:nncf_zz}
    F^{(L)}_{zz}(\vec{x},\vec{x}')= {}&
    \sum_{\ell=1}^{L} \sum_{m=-\ell}^{m=-\ell}
    A_{\ell}\bigg(\frac{a^2}{rd}\bigg)^{\ell +2}\bigg[(\ell+1)Y^*_{\ell m}(0,0)\bigg]\nonumber\\
    {}&\bigg[\cos\theta(\ell +1)Y_{\ell m}(\theta,0)+\sin\theta\frac{\partial}{\partial\theta}Y_{\ell m}(\theta,0)\bigg]
\end{align}
Plotting Eq.\ref{eqn:nncf_zz} for values of $d=2a$ and $d=5a$ gives us Figs. \ref{fig:nonlocal_bzbz_m=2_0} and \ref{fig:nonlocal_bzbz_m=5_0}.  the $zz$ entry for the nonlocal NCF is the $z$-component of the induced field for a fictitious dipole at $\vec{x^\prime}=(0,0,d)$ pointing in the $z$-direction.  This induces a dipole on the sphere that points in the $z$-direction.  This explains why the we see a highly positive correlation near the $z$-axis and a highly negative one along the $x$-axis.

 In Fig.\ref{fig:nonlocal_bzbz_m=2_0}, in which $d=2a$, the higher multipoles make a very significant contribution to the NCF and an interesting asymmetric pattern emerges. When $F^{(L)}_{zz}(\vec{x},\vec{x}')$ is calculated up to the even harmonics ($l=2,4$), the positively valued region in the negative $z$-axis becomes smaller. This is caused by the effects of odd and even harmonics in affecting the symmetry of the function under reflection in the $x-y$ plane. The even harmonics are asymmetric and the odd harmonic are symmetric under this reflection operation, which gives the periodic behavior as a function of $L$.  At the level of resolution of the figures (a few percent), the function has converged at around $L=4-5$.  In contrast, Fig.\ref{fig:nonlocal_bzbz_m=5_0} in which $d=5a$ we see that the higher order multipole terms do not contribute much to the field.  This is expected since for large $d$ the fictitious field is nearly uniform at the sphere and higher-order dipoles are not induced. Convergence happens at about $L=1-2$.  
 

\subsection{$F^{(L)}_{xx}(\vec{x},\vec{x}')$ for various limits L on the sum}
\label{subsec:nncf_xx}
\begin{figure}[hbt!]
    \centering
    \includegraphics[width=\columnwidth]{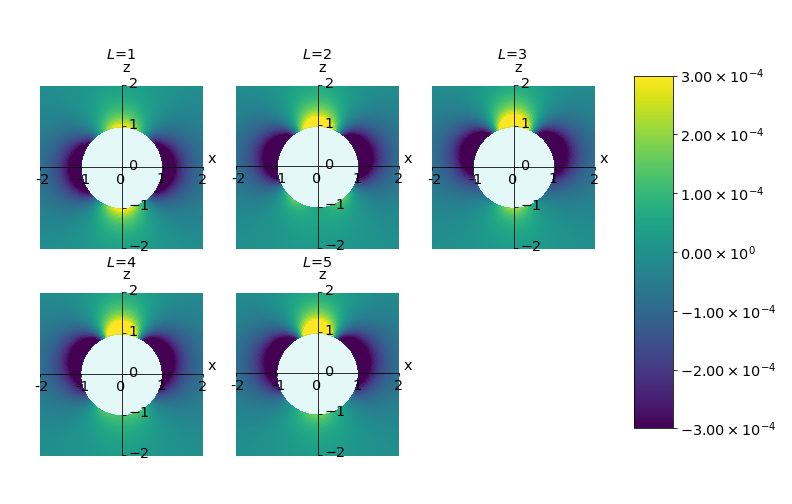}
   \caption{Dimensionless nonlocal noise correlation function $F^{(L)}_{xx}(\vec{x},\vec{x}')$.  Here $\vec{x}'=(0,0,2a)$ (at top center of the plots) is fixed and the plot is as a function of $\vec{x}$.  The unit of distance is $a$, the radius of the sphere.  Shown is the function obtained for $L=1$ to $L=5$, where $L$ is the number of multipoles included in the sum. Deep-blue color indicates regions of negative correlation and yellow color shows regions of positive correlation.}
    \label{fig:nonlocal_bxbx_m=2_0}
\end{figure}

\begin{figure}[hbt!]
    \centering
    \includegraphics[width=\columnwidth]{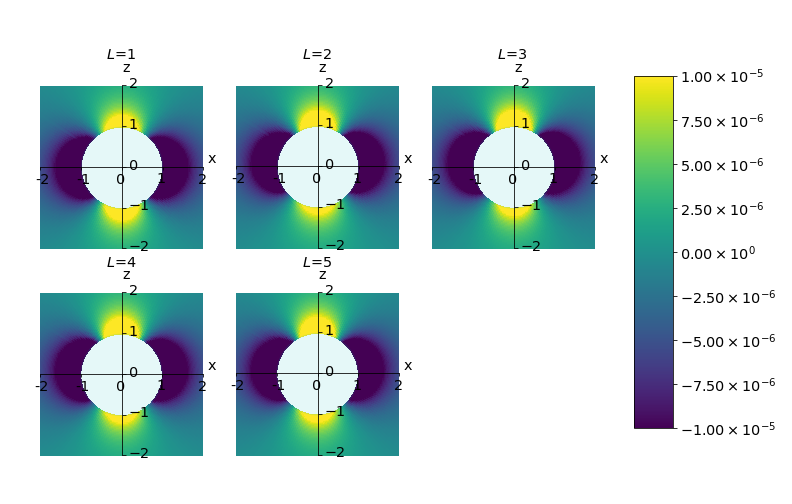}
    \caption{Dimensionless nonlocal noise correlation function $F^{(L)}_{xx}(\vec{x},\vec{x}')$.  Here $\vec{x}'=(0,0,5a)$ (off the plots) is fixed and the plot is as a function of $\vec{x}$.  The unit of distance is $a$, the radius of the sphere.  Shown is the function obtained for $L=1$ to $L=5$, where $L$ is the number of multipoles included in the sum. Deep-blue color indicates regions of negative correlation and yellow color shows regions of positive correlation.}
    \label{fig:nonlocal_bxbx_m=5_0}
\end{figure}

Solving equation (\ref{eqn:nonlocal_xx}), we get
\begin{align}
    \label{eqn:nncf_xx}
    F^{(L)}_{xx}(\vec{x},\vec{x}')= {}& \sum_{\ell=1}^{L} \sum_{m=-\ell}^{m=-\ell}
    A_{\ell}\bigg(\frac{a}{rd}\bigg)^{\ell +2}(d)^{-\ell -2}\\
    {}&\bigg[-\frac{\partial}{\partial\theta}Y^*_{\ell m}(0,0)\bigg]\nonumber\\
    {}&\bigg[\sin\theta(\ell +1)Y_{\ell m}(\theta,0)-\cos\theta\frac{\partial}{\partial\theta}Y_{\ell m}(\theta,0)\bigg]
\end{align}
Plotting Eq. (\ref{eqn:nncf_xx}) for various values of $d$ gives us Fig. \ref{fig:nonlocal_bxbx_m=2_0} and \ref{fig:nonlocal_bxbx_m=5_0}. The fictitious dipole at $\vec{x^\prime}=(d,0,0)$ is now pointing in the $x$-direction induces a dipole on the sphere pointing in the $-x$ direction situated at the origin. Therefore, the NCF is highly negative along the $x$-axis and positive along the $z$-axis. Similar asymmetry and symmetry patterns are observed for the odd and even harmonics of the function. The asymmetry is observed along the $x$-axis because the fictitious dipole is near the north pole of the sphere. It causes more induced current flowing in the northern hemisphere than the southern hemisphere.  The even-odd pattern in $L$ is even more pronounced than in the previous case. As a result the convergence is even slower, happening around $L=5-6$   Similar to the figures in subsection \ref{subsec:nncf_zz}, the higher order terms do not cause very significant change to the field as observed in Fig.\ref{fig:nonlocal_bzbz_m=5_0}, and the function converges at $L=2-3$

\subsection{$F^{(L)}_{xz}(\vec{x},\vec{x}')$ for various limits $L$ on the sum}
Solving equation (\ref{eqn:nonlocal_xz}), we get
\begin{align}
    \label{eqn:nnfc_xz}
    \vec{B}^x_f(\vec{x},\vec{x}') = {}& \sum_{\ell=1}^{L} \sum_{m=-\ell}^{m=-\ell}
    A_{\ell}\bigg(\frac{a}{rd}\bigg)^{\ell +2}\bigg[(l+1)Y^*_{lm}(0,0)\bigg]\nonumber\\
    {}&\bigg[\sin\theta(l+1)Y_{lm}(\theta,0)-\cos\theta\frac{\partial}{\partial\theta}Y_{lm}(\theta,0)\bigg]
\end{align}

\begin{figure}[hbt!]
    \centering
    \includegraphics[width=\columnwidth]{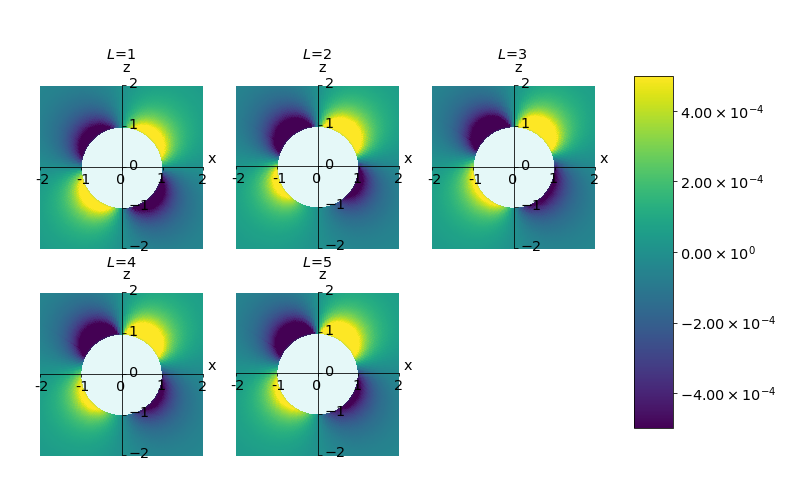}
    \caption{Dimensionless nonlocal noise correlation function $F^{(L)}_{xz}(\vec{x},\vec{x}')$.  Here $\vec{x}'=(0,0,2a)$ (top center of the plots) is fixed and the plot is as a function of $\vec{x}$.  The unit of distance is $a$, the radius of the sphere.  Shown is the function obtained for $L=1$ to $L=5$, where $L$ is the number of multipoles included in the sum. Deep-blue color indicates regions of negative correlation and yellow color shows regions of positive correlation.}
    \label{fig:nonlocal_bxbz_m=2_0}
\end{figure}

\begin{figure}[hbt!]
    \centering
    \includegraphics[width=\columnwidth] {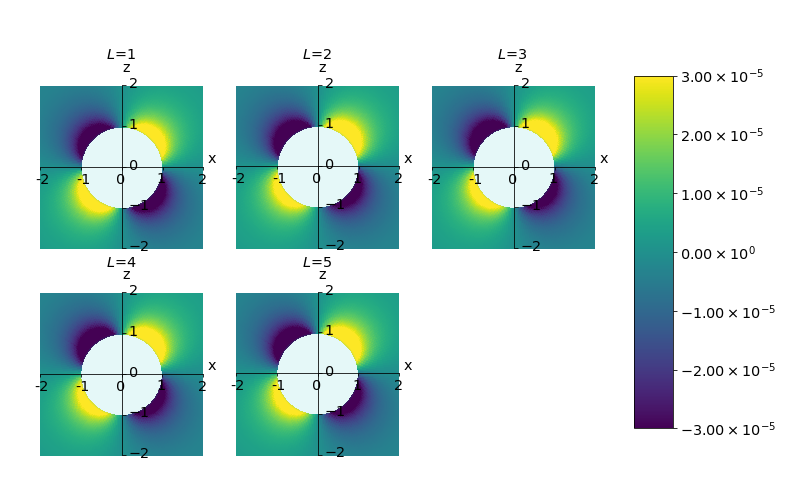}
   \caption{Dimensionless nonlocal noise correlation function $F^{(L)}_{xz}(\vec{x},\vec{x}')$.  Here $\vec{x}'=(0,0,5a)$ (off the plots) is fixed and the plot is as a function of $\vec{x}$.  The unit of distance is $a$, the radius of the sphere.  Shown is the function obtained for $L=1$ to $L=5$, where $L$ is the number of multipoles included in the sum. Deep-blue color indicates regions of negative correlation and yellow color shows regions of positive correlation.}
    \label{fig:nonlocal_bxbz_m=5_0}
\end{figure}

This is the only off-diagonal entry in the non-local NCF that we will investigate.  We must now imagine a fictitious dipole that points in the $z$-direction and we observe the $x$-component of the induced field.   
Plotting Eq. (\ref{eqn:nnfc_xz}) for different values of $d$ gives us Figs. \ref{fig:nonlocal_bxbz_m=2_0} and \ref{fig:nonlocal_bxbz_m=5_0}. Similar to subsection \ref{subsec:nncf_zz}, the fictitious dipole induces a dipole on the sphere pointing in the $+z$-direction at the origin. However, the function $F_{xz}$ maps the $x$ component of the noise field. The magnetic field created by this induced dipole on the sphere goes outward from the north pole, going around the equator of the sphere, and comes back inward into the south pole. Therefore, one would expect the $x$-component to be positive in the $x>0,z>0$ and $x<0,z<0$ quadrants and negative in the $x>0,z<0$ and $x<0,z>0$ quadrants, as is indeed seen in the plots. The asymmetry of the higher order terms is the same as has been explained above in subsection \ref{subsec:nncf_xx}.

\subsection{Implications for Quantum Computing}
\label{subsec:nonlocal}

The recommendations for quantum computer designers that come from the angular pattern of the nonlocal NCF are slightly more subtle than those for the local NCF, but not different in kind.   Above, we gave a hardware recommendation.  Here, we give a software example.  

Again there is a single metallic object at the origin but now we have two spin qubits, one at $\vec{x}$ and one at $\vec{x}'$.  The nonlocal NCF $F_{ij}(\vec{x},\vec{x}')$ couples to an operator $\sigma^{\vec{x}}_i \sigma^{\vec{x}'}_j $, where $\sigma_i^{\vec{x}}$ is the Pauli matrix that acts on a spin at $\vec{x}$.  The noise creates decoherence on \textit{nonlocal} quantum amplitudes. 

Here is an example that is useful for algorithm design.  Let us say that the qubits are at positions where the $zz$ entry of $F_{ij}(\vec{x},\vec{x}')$ is large compared to other entries. A glance at Figs.\ref{fig:nonlocal_bzbz_m=2_0} through \ref{fig:nonlocal_bxbx_m=5_0} shows that this is the case, for example, when both qubits are on the $z$-axis but on opposite sides of the object.  We can protect the qubits from the noise by working in a decoherence-free subspace.  Let us choose the computational basis $ \{| 0 \rangle, |1 \rangle \} $ of eigenstates of $\sigma_z$.  The Bell state $ (1/\sqrt{2}) ( |00 \rangle + |11 \rangle)$ is decoherence-free if only $zz$ noise is present.  In particular, the relative phase of the  $  |00 \rangle $ and $ |11 \rangle $ states is preserved.  This is not the case for the relative phase of $ ( |00 \rangle $ and $ |01 \rangle)$ states.  Hence we wish to operate in the  $\{( |00 \rangle , |11 \rangle) \}$ subspace and this can be done by careful design of the quantum circuit that implements a quantum algorithm.  

Other examples are easy to construct. 

\section{Conclusion}
\label{sec:conclusion}
Controllable qubits with long decoherence time are desirable in quantum computing. This means that the existence of metallic elements of qubit devices are a double-edged sword. On one hand, they are needed to interact with and thereby control the qubits. On the other hand, the fluctuations of currents and charges in these metallic objects create a noise field that disturbs and decoheres the qubits. 
The results in this paper allowed us to visually represent the noise field in the different spatial points of qubit arrays when metallic objects are present.  We  gave examples of how this visualization can aid in the design of both hardware and software for a quantum computer.

From a more formal and mathematical point of view, we gave a solution in closed form for the magnetic EWJN for an object, or set of objects in the limits appropriate for a nano-device.  Its simplicity means that it can be used as the basis for numerical calculations of the NCF for real devices.  We also calculated, for the first time, the multipole expansion for the NCF of a spherical metallic device.  The characteristics of the solution illustrated the general principle of magnetic EWJN from localized objects.

\begin{acknowledgements}
We thank M.G. Vavilov and V.N. Premakumar for useful discussions.  This research was sponsored in part by the Army Research Office (ARO) under Grant Number W911NF-17-1-0274.The views and conclusions contained in this document are those of the authors and should not be interpreted as representing the official policies, either expressed or implied, of the Army Research Office (ARO), or the U.S. Government. The U.S. Government is authorized to reproduce and distribute reprints for Government purposes notwithstanding any copyright notation herein.
\end{acknowledgements}

\bibliographystyle{apsrev4-1}
\bibliography{aapmsamp}

\end{document}